# Demonstrating the Potential of Adaptive LMS Filtering on FPGA-Based Qubit Control Platforms for Improved Qubit Readout in 2D and 3D Quantum Processing Units

Hans Johnson [1,2,†], Nicholas Bornman [2], Taeyoon Kim [2,3], David Van Zanten [2], Silvia Zorzetti [2], and Jafar Saniie [1]

[1] Embedded Computing and Signal Processing (ECASP) Research Laboratory,
Department of Electrical and Computer Engineering,
Illinois Institute of Technology, Chicago, IL 60616, U.S.A.

[2] Superconducting Quantum Materials and Systems (SQMS) Center,
Fermi National Accelerator Laboratory, Batavia, IL 60510, U.S.A.

[3] Center for Applied Physics and Superconducting Technologies (CAPST),
Department of Physics and Astronomy,
Northwestern University, Evanston, IL 60208, U.S.A.

[†] hjohnson1@hawk.iit.edu

*Abstract*—Advancements in quantum computing underscore the critical need for sophisticated qubit readout techniques to accurately discern quantum states. This abstract presents our research intended for optimizing readout pulse fidelity for 2D and 3D Quantum Processing Units (QPUs), the latter coupled with Superconducting Radio Frequency (SRF) cavities. Focusing specifically on the application of the Least Mean Squares (LMS) adaptive filtering algorithm, we explore its integration into the FPGA-based control systems to enhance the accuracy and efficiency of qubit state detection by improving Signal-to-Noise Ratio (SNR). Implementing the LMS algorithm on the Zynq UltraScale+ RFSoC Gen 3 devices (RFSoC 4x2 FPGA and ZCU216 FPGA) using the Quantum Instrumentation Control Kit (QICK) open-source platform, we aim to dynamically test and adjust the filtering parameters in real-time to characterize and adapt to the noise profile presented in quantum computing readout signals. Our preliminary results demonstrate the LMS filter's capability to maintain high readout accuracy while efficiently managing FPGA resources. These findings are expected to contribute to developing more reliable and scalable quantum computing architectures, highlighting the pivotal role of adaptive signal processing in quantum technology advancements.

## I. Introduction

Quantum computing is transforming the landscape of computational problem-solving and will equip us with the ability to tackle challenges that defy some of the current capabilities of classical computers. At the core of this transformation are quantum bits, or qubits, which hold the promise of exponential speed-ups in computing by leveraging their ability to exist in multiple states simultaneously. It is this ability to exploit the quantum mechanical property of superposition that gives quantum computers their extraordinary power. Yet, realizing the full promise of quantum computing demands overcoming significant obstacles in experimental physics, materials, and engineering. Qubits are notoriously sensitive as their quantum states are easily disrupted by environmental noise, leading to decoherence. To combat this, researchers are innovating with error-correcting techniques [1-3], developing new materials [4-6], and refining control systems to maintain qubit stability [7-9].

At the forefront of those developments, superconducting qubits are making waves, particularly within the framework of 2D and 3D Quantum Processing Units (QPUs). At the Superconducting Quantum Materials and Systems (SQMS) Center at Fermilab, the approach is to utilize and test fabricated qubits (2D QPU) as well as couple these qubits to Superconducting Radio Frequency (SRF) cavities (3D QPU) as resonators for the physical qubits. This research into 2D and 3D QPU aims to curtail noise and bolster qubit coherence [4]. Here, precise Radio Frequency (RF) pulses are key; they delicately interact with qubits by dictating their state with extreme precision using control pulses, and carrying back computational results with readout pulses.

This is where Field-Programmable Gate Arrays (FPGAs) come into play. FPGAs offer high-speed processing and extremely low latency necessary for quantum control and readout. Although control and readout equipment can also be done through commercial systems – such as an RF stack of Arbitrary Waveform Generators (AWGs) – these conventional control stacks are, on average, ten times as expensive as an FPGA platform which shows no difference in degradation of coherence to some commercial systems [10]. The Quantum Instrumentation Control Kit (QICK) from Fermilab's Scientific Computing Division is an open-source and hardware-agnostic project that is meant to run on Xilinx's UltraScale+ Radio Frequency System-on-Chip (RFSoC) FPGA line, and is a testament to the potential and integration of FPGAs in quantum systems. Though effective, there is room for optimization, especially regarding the Signal-to-Noise Ratio (SNR) and readout pulse fidelity for both 2D and 3D QPU experiments. Though there exist other FPGA platforms for controlling qubits [11-12], none specifically address optimizations via filtering for control or readout pulse fidelity as a key focus of the research.

This research seeks to address this area by deploying an adaptive filtering technique to clarify the readout signals that are the basis of quantum state discrimination and qubit assessment. Using an RFSoC FPGA-based control board, we built an adaptive Least Means Squared (LMS) filter for the sake of dynamically fine-tuning the readout and enhancing signal fidelity amidst the inherent noise. This work focuses on the development of an adaptive filter to be deployed onto open-source control platforms, such as QICK, and presents preliminary results of its development. This filter is meant to be placed at the intersection of classical signal processing and quantum information, showing how refined control and readout can bolster quantum computation and bring us closer to harnessing the true power of qubits for practical applications.

Section II delves into the system design in terms of hardware used for both our quantum systems and FPGA-based control hardware, as well as a more detailed look into the methodology behind firmware and software development for our quantum computing platforms. Section III discusses the exact implementation of the adaptive filter into a demonstration that proves the efficacy of an LMS filter design on FPGA for signals that have near identical parameters to readout signals used in qubit experiments. Section IV shows some of the preliminary results given by the demo and describes how this fits into the context of qubit experiments for future integration. Concluding remarks and future work is given in Section V.

## II. SYSTEM DESIGN

### A. Hardware

The control board used for this project is the RFSoC 4x2, which is a board released by Real Digital built around AMD Xilinx's Zynq UltraScale+ RFSoC ZU48DR device [13]. This FPGA chip is a part of Xilinx's Gen3 RFSoC line, which are primarily used in quantum system FPGA-based controls like the ZCU216 and RFSoC 4x2 boards, as well as the Gen1 RFSoC line such as the ZCU111 board [10, 14-15]. The RFSoC line is widely adopted because it integrates multiple high-speed ADCs and DACs combined with a multi-core ARM processing system and FPGA in a single System-on-Chip (SoC), which eliminates the need for external data converters and enables ultra-accurate synchronization and channel timing. The RFSoC 4x2 features the following RF peripheral specifications, and a picture of the board can be found in Fig. 1:

- Four accessible 14-bit 5 GSPS RF-ADCs
- Two accessible 14-bit 9.85 GSPS RF-DACs

Though describing quantum computing hardware is beyond the scope of this research, among some of the resources available for 2D and 3D QPU is a comprehensive description for research at SQMS found in [16].

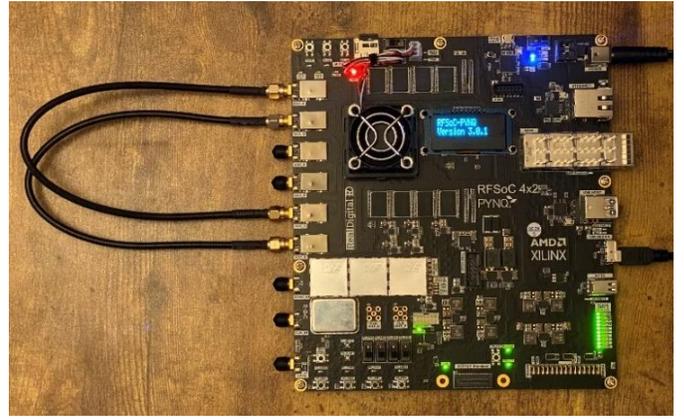

Fig. 1. Xilinx UltraScale+ RFSoC 4x2 Board with Physical Loopbacks connected with two RF SMA coaxial cables.

### B. Firmware

In FPGA-based systems, firmware Intellectual Property (IP) is a functional block used to manage hardware efficiently. They are typically written in Hardware Description Languages such as VHDL or Verilog. These IPs are modular pieces of code that handle specific tasks such as generating control signals, reading sensor outputs, or interfacing with external hardware or software devices. Once integrated into an FPGA's firmware, these IPs interact directly with the FPGA's programmable logic to control data flow and signal processing in real time.

In quantum computing architectures, particularly those involving superconducting qubits, control and readout signals are crucial for qubit manipulation and measurement. These signals are typically managed through precise firmware that dictates the timing, amplitude, and phase of pulses sent to the qubits. There are two types of signals in quantum systems generated in control hardware: control and readout signals.

Control signals are used to perform quantum operations on qubits. For superconducting qubits like transmon qubits, microwave pulses tuned to the qubit's resonance frequency (determined by the Josephson junctions and the capacitance in the circuit) are used. These pulses are generated following the Jaynes-Cummings model, where a two-level system (qubit) interacts with a quantized harmonic oscillator (resonator) in a controlled manner to achieve specific quantum states [17-19].

Readout signals are used to measure the state of the qubits after qubit manipulation. This is often achieved through a process called Quantum Non-Demolition (QND) measurement using a microwave resonator coupled to each qubit. The resonator frequency shifts depending on the qubit state, and by probing this resonator, one can infer the qubit's state without destroying its quantum information [20-21].

### C. Software

In the Python Productivity for Zynq (PYNQ) framework, firmware can be managed with a software overlay for control and readout systems. In short, a user can use Software Defined Radio (SDR) to design control and readout signals as well as manage hardware peripherals at a high-level to control variables such as waveform parameters, sampling rates, clock speeds, and more. This allows for rapid prototyping and testing for not only FPGA developers, but users familiar with Python [22].

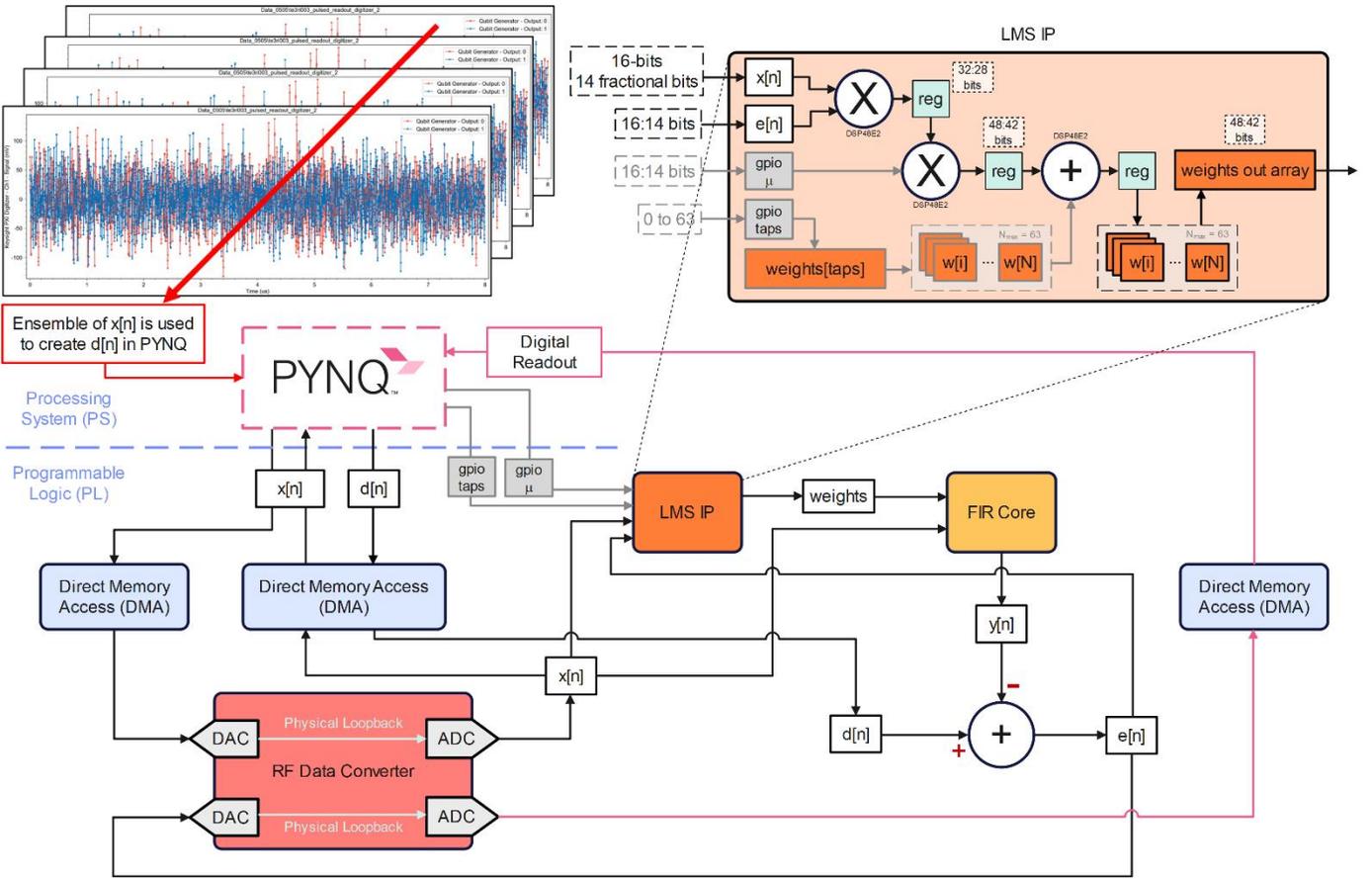

Fig. 2. Block Diagram of LMS Demonstration Project on RFSoC 4x2: A noisy 'x' signal was characterized by single-shot readout pulse data for 3D QPU experiments for qubit measurement. The 'x' signals for the demo were built using Python in the PYNQ overlay and sent from the Processing System (PS) to the Programmable Logic (PL) via Direct Memory Access (DMA) IPs. The 'x' signal was passed to the RF Data Converter (RFDC) IP to simulate an actual readout signal chain for qubit measurements, and the physical DACs and ADCs were connected in loopback with SMA cables as seen in Fig. 1. The LMS algorithm is implemented within the LMS IP, and a description of the LMS equations along with the signal definitions can be found in Section III.

## III. Implementation

The demonstration of the LMS algorithm in firmware was built in the context of QICK and meant to simulate what will occur when this IP is inserted into a firmware platform for quantum computing readout. The purpose of the demo is to prove the efficacy of an LMS implementation on the RFSoC platform applied to readout signals after they interact with the qubit resonator and are sent back into the control system via ADC. This was simulated by creating a series of sample pulses characterized by 3D QPU single-shot time trace readout pulse data. Typically, the actual control and readout pulses for both 2D and 3D QPU operate in the GHz regime due to the superconducting properties of the qubit and related hardware in the circuit chain. Superconducting materials, which are used to construct qubits and their associated resonators, have a property called the superconducting gap. This gap represents the energy difference required to break Cooper pairs (pairs of electrons that are bound together at low temperatures) and create quasiparticles. Since the superconducting gap corresponds to an energy equivalent in the microwave GHz range, this conveniently sets the operational frequencies of superconducting qubits and resonators into the GHz spectrum [23].

Although these pulses are typically around 2-10 GHz depending on the qubit and resonator design, when readout pulses are received back into the control system they are subject to digital downconversion to manage and analyze these high-frequency signals effectively. After the data enters the digital domain, digital down-conversion (DDC) is applied. Generally, in DSP this allows for more precise control over the mixing and filtering processes and enhanced noise performance through steeper roll-offs and better rejection of out-of-band noise for digital filters. It is also much easier for hardware resources to filter in the MHz regime than GHz. This is especially important in quantum computing because exact manipulation of signals determines the success of experimental outcomes.

Therefore, these down-converted signals are the signals we want to target for filtering. The mock-readout signals after interacting with the qubit resonator were generated to be 30 MHz pulses over 8 μs, which match the profiles for some of the readout signals used in experiments at SQMS. The design is meant to be a testbed for exploring different frequencies, pulse periods, and noise profiles of signals along with varying learning rates and taps used by the adaptive filter, but testing was aligned closely with what we expect to see for actual qubit experiments. A block diagram can be found in Fig. 2.

The parameters of the LMS Filter demo highlighted in Fig. 2 can be found below:

- n: Sample index
- i: Tap index within the filter's impulse response
- x[n]: Noisy input signal
- d[n]: Desired/reference signal
- w[i]: Weights = adjustable weights of LMS
- y[n]: Output signal (estimated signal or noise estimation) at time index [n]
- e[n]: error signal (difference between desired and output signals) at time index [n]
- μ: Learning rate of LMS algorithm
- N: number of taps in the filter, dictating the number of past input samples used by the filter

The mathematical operations for signal processing defined in equations (1), (2), and (3) were implemented in custom IP blocks using VHDL, while equation (4) was executed within the PYNQ overlay through DMA access. This integration was achieved in the firmware, connecting to the processing system via the AXI protocol. The weights update is given below in (1) and was handled by the custom LMS IP:

$$w[i+1] = w[i] + \mu * e[n] * x[n-i] \qquad (1)$$

The output signal calculation, 'y', from the FIR filter is given below in (2), and was calculated using the FIR Core from Xilinx's default IP library and reloading the calculated weights into the FIR core's coefficients for the 'x' samples that pass through the IP:

$$y[n] = \sum_{i=0}^{N-1} w[i] * x[n-i] \qquad (2)$$

The error signal, 'e', is calculated by a custom RTL module IP that takes the cumulative average, 'd' (generated from 'x'), and subtracts is from the output from the FIR filter, 'y'. This module also synchronizes the samples from 'd' and 'y' using the TLAST call in the AXI4-Stream handshake protocol so the two do not become misaligned. The equation is given by (3):

$$e[n] = d[n] - y[n] \qquad (3)$$

Finally, the desired signal is calculated through a cumulative average after PYNQ acquires 'x' back through the DMA to simulate how it would see the noisy readout signal for a qubit readout experiment. This equation takes each sample of x[n] and calculates an average, 'd' for each pulse. Each averaged sample for d[n] is outputted from the DMA once the last sample for 'x' is reached, and the calculation is given by (4):

$$d_{new}[n] = \frac{(d_{old}[n] * i) + x_{new}[n]}{i+1} \qquad (4)$$

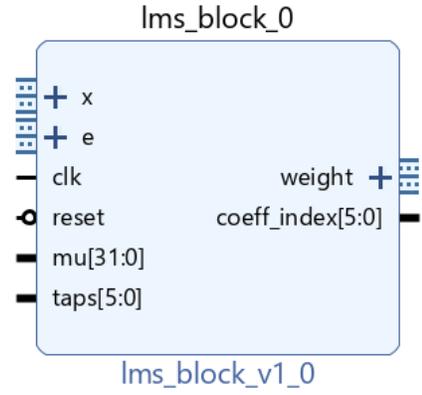

Fig. 3. LMS IP: Packaged IP for the LMS Block that can be easily invoked using the IP integrator in AMD Xilinx Vivado.

The packaged IP of the LMS block can be found in Fig. 3. This project can also be built without using project mode in Vivado. A table of the resource utilization can be found below in Table I for the demo project as a whole. This closely represents what the final packaged IP for the LMS filter will look like as the three DMAs and FIR Core will need to be packaged together for a final implementation to insert into FPGA firmware, such as the QICK platform.

The LMS IP deployed on the RFSoC 4x2 platform utilizes 10 DSP48E2 slices for high-speed arithmetic operations essential for real-time adaptive filtering, accounting for just 0.23% of the total DSP resources available. The functional IPs for data flow were clocked at 491.52 MHz, while the register IPs were invoked with a clock of 100 MHz. The efficient use of DSPs underscores the IP's optimization for performance without overwhelming the FPGA's capabilities. The pipelined architecture of the VHDL code involves multiple stages: multiplication of input and error signals, intermediate storage, further multiplication with the adaptation step, and accumulation with clipping, ensuring seamless data flow and latency reduction. The pipeline can be observed at the top of Fig. 2 for the description of the LMS IP. These stages are crucial for dynamic updating of filter coefficients in response to changing signal conditions. Additionally, the design incorporates 21,964 LUTs and 33,992 FFs, managing logic and state operations, alongside 133 BRAMs for extensive data and coefficient storage, indicative of the implementation's sophisticated memory management.

TABLE I: RESOURCE UTILIZATION FOR LMS DEMO

| Resource | Utilization | Available | Utilization (%) |
|---|---|---|---|
| LUT | 21,964 | 425,280 | 5.16% |
| LUTRAM | 3,529 | 213,600 | 1.65% |
| FF | 33,992 | 850,560 | 4.00% |
| BRAM | 133 | 1080 | 12.31% |
| DSP | 10 | 42,732 | 0.23% |
| BUFG | 7 | 696 | 1.01% |
| MMCM | 1 | 8 | 12.5% |

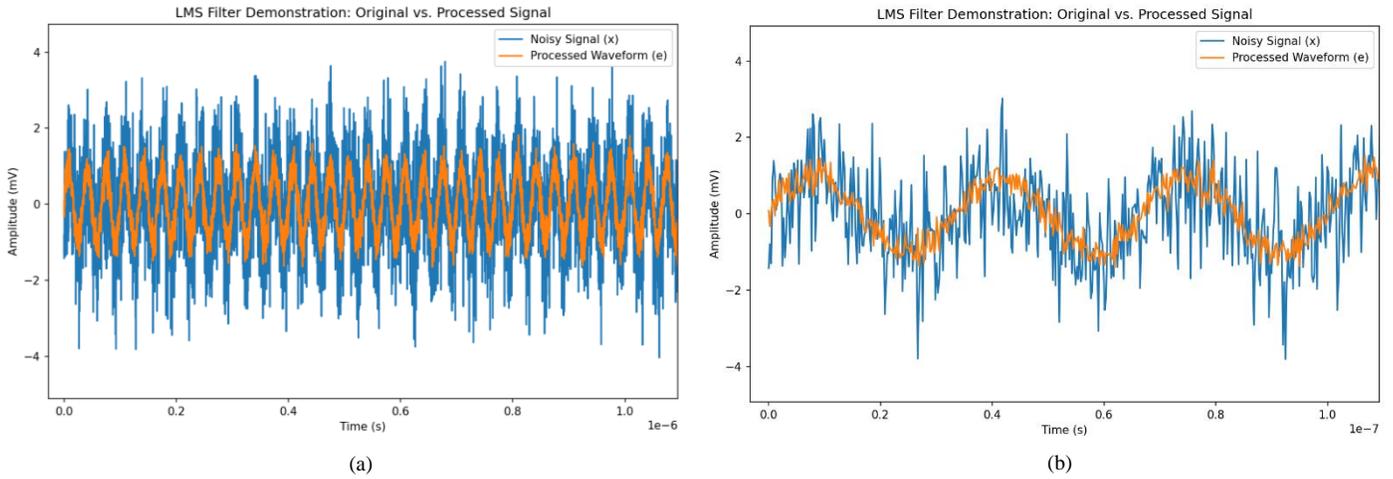

Fig. 4. LMS IP Demonstration of Noisy Signal (x) and LMS Processed Waveform (e): a) Zoomed in window of 1 µs. b) Zoomed in window of 0.1 µs. Waveform characterization is a 30 MHz, 8 µs pulse with a white-noise standard deviation of 1. The learning rate of the LMS filter was set to 0.0006 and the number of taps was 64. The filter was allowed to execute over 10 pulses for this demo, and the 10th pulse/iteration for both 'x' and 'e' are shown on the graph.

## IV. RESULTS

In the demonstration of the LMS filter, the processed waveform 'e' shows significant noise attenuation from the original noisy signal 'x', across a 30 MHz, 8 µs pulse waveform. This outcome was achieved with the LMS filter's learning rate set at 0.0006 and the filter employing 64 taps. Notably, the waveforms depicted in Figure 4 for both the 1 µs (a) and the 0.1 µs (b) windows, which represent the 10th pulse/iteration, exhibit a well-maintained alignment, indicating no discernible latency between the processed and the input noisy signals.

The LMS filter's noise reduction capabilities are clearly evidenced in the signal processing results. Noise characterization of the original signal 'x' reveals a consistent white noise floor across the 30 MHz bandwidth, as expected from a 30 MHz, 8 µs pulse with the defined noise parameters. The LMS filter, exhibits robust noise cancellation in the processed waveform 'e', even after just 10 pulses. This suggests that the LMS algorithm is capable of effectively converging.

The absence of observable latency is intriguing, as one might expect the iterative adaptive process of the LMS algorithm to introduce a delay. Several factors contribute to this outcome, but the main one is the demo was built to treat the series of input pulses as a continuous stream of data using the handshake protocols. Another is the physical loopback connection made with short coaxial cables minimizes the propagation time between the DAC and the ADC, potentially reducing the observable latency to below our measurement resolution. Additionally, the synchronization IP employed in the system effectively aligns the input and output data streams, which may negate the processing delay introduced by the FIR filtering within the LMS algorithm. Another consideration is the pipelined design of the VHDL implementation, which is optimized for high throughput and may allow real-time updating of coefficients with negligible impact on the signal path latency. Future investigations should aim to quantify the system latency through targeted experiments designed to separate the contributions of hardware synchronization and algorithmic processing.

After 10 iterations, evident in the 10th pulse displayed for both 'x' and 'e', the filter significantly mitigates high-frequency noise while preserving the signal envelope—critical for quantum signal readouts. This can be scaled up for many more pulses. The improved SNR in the output signal demonstrates the LMS filter's potential to enhance readout fidelity in quantum computing applications, affirming the value of real-time adaptive filtering as adaptive coefficient adjustments keep pace with incoming signal variations. Results were similar for different standard deviations of noise, but frequency and pulse period were kept the same for all tests.

This application is especially important because the current methodology for extracting a clean readout signal is to average 10,000 to 100,000 single-shot time traces to get a signal that a researcher can discern between the qubit's ground and excited state. Though the formation of the desired signal, 'd', follows this ensemble method, the filter is designed to characterize the noise profile based on this ensemble averaging and adjust the coefficients as the signal passes through the system. This active filtering utilizes this ensemble method on a per-pulse basis, and allows the filtered signal to converge faster to the noiseless signal than any sort of ensemble averaging in post-processing would. This means that as the synchronizer IP attempts to unite the desired signal 'd' with the filtered signal 'y', the noise will be filtered out better and better for each pulse that passes through the filter. Though the next step is to apply actual streaming qubit data to the filter, the evidence here suggests we can achieve noise cancellation during an active experiment.

Once the demo application is completely refined on the 4x2 platform, the LMS application will be moved on to the ZCU216 to test with real QPU. So far, the precise noise attenuation observed in the LMS-filtered waveforms underscores the filter's proficiency in real-time signal clarification, essential for the fidelity of quantum readout. This suggests that the integration of such a filter could notably reduce the need for extensive ensemble averaging, and streamlining data analysis in quantum experiments.

## V. Conclusion and Future Work

The demonstrated LMS filter, with its adaptive noise cancellation capabilities, presents a pivotal use case in qubit characterization experiments. It is instrumental in measuring parameters such as T1 (energy relaxation times), T2 (dephasing times), and coherence times, which are foundational for the performance assessment of qubits. By enhancing signal fidelity, the LMS filter facilitates more accurate discrimination of quantum states, contributing to the precision of quantum logic gates and readout processes. Additionally, its real-time processing ability can be leveraged in quantum error correction protocols to discern and correct errors as they occur, thus potentially increasing the robustness and reliability of quantum computing architectures.

Though more testing needs to be completed, the demo proves the efficacy of implementing an adaptive filter algorithm for ANC-based quantum computing readout. Once implemented in QICK for 2D QPU experiments and in the readout signal chain for 3D QPU experiments, we can quantify the improvement of SNR on actual qubit data, which will provide significantly valuable insight and data to the use case of adaptive filters and FPGAs in the control stack. It is vital to also observe how adjusting the learning rate and taps will affect the transient and steady-state behavior of the filter across different qubits. As we extend the application of the LMS filter to more complex quantum systems, its integration promises not only to refine the current paradigms of qubit readout fidelity but also to contribute to the overarching goal of realizing more stable and scalable quantum computing technologies.

## Acknowledgments

This material is based on work supported by the U.S. Department of Energy, Office of Science, National Quantum Information Science Research Centers, Fermilab's Superconducting Quantum Materials and Systems Center (SQMS), and Fermilab's Scientific Computing Division under contract No. DE-AC02-07CH11359. This material is also based upon work supported by the U.S. Department of Energy, Office of Science, Office of Workforce Development for Teachers and Scientists, and Office of Science Graduate Student Research (SCGSR) program. The SCGSR program is administered by the Oak Ridge Institute for Science and Education for the DOE under contract number DE-SC0014664.

## References

[1] Google Quantum AI, "Suppressing quantum errors by scaling a surface code logical qubit", Nature 614, 676–681 (2023). https://doi.org/10.1038/s41586-022-05434-1

[2] A. Jayashankar, P. Mandayam, "Quantum Error Correction: Noise-Adapted Techniques and Applications", J Indian Inst Sci 103, 497–512 (2023). https://doi.org/10.1007/s41745-022-00332-x.

[3] D. Cruz, F. A. Monteiro, B. C. Coutinho, "Quantum Error Correction Via Noise Guessing Decoding", IEEE Access, vol. 11, pp. 119446-119461, 2023, https://doi.org/10.1109/ACCESS.2023.3327214.

[4] M. Bal, et al., "Systematic Improvements in Transmon Qubit Coherence Enabled by Niobium Surface Encapsulation", arXiv preprint Jan. 2024 https://doi.org/10.48550/arXiv.2304.13257. Accepted for publication in Nature Partner Journal Quantum Information.

[5] J. Lee, et al., "Stress-induced structural changes in superconducting NB Thin films." Physical Review Materials, vol. 7, no. 6, 27 June 2023, https://doi.org/10.1103/physrevmaterials.7.l063201.

[6] D. Pappas, et al., "Alternating Bias Assisted Annealing of Amorphous Oxide Tunnel Junctions", arXiv preprint February 2024, https://doi.org/10.48550/arXiv.2401.07415. Under review at Nature Portfolio.

[7] C. Ding, et al., "Experimental advances with the QICK (Quantum Instrumentation Control Kit) for superconducting quantum hardware." Physical Review Research, vol. 6, no. 1, 20 Mar. 2024, https://doi.org/10.1103/physrevresearch.6.013305.

[8] L. Qiu, R. Sahu, W. Hease, G. Arnold, J. M. Fink, "Coherent optical control of a superconducting microwave cavity via electro-optical dynamical back-action", Nature Communications 14, 3784 (2023). https://doi.org/10.1038/s41467-023-39493-3.

[9] R. Acharya, et al., "Multiplexed superconducting qubit control at millikelvin temperatures with a low-power cryo-CMOS multiplexer", Nature Electronics 6, 900–909 (2023). https://doi.org/10.1038/s41928-023-01033-8.

[10] L. Stefanazzi, et al. "The QICK (Quantum Instrumentation Control Kit): Readout and control for qubits and detectors." *Review of Scientific Instruments*, vol. 93, no. 4, 1 Apr. 2022, https://doi.org/10.1063/5.0076249.

[11] Y. Xu et al., "QubiC: An Open-Source FPGA-Based Control and Measurement System for Superconducting Quantum Information Processors," in IEEE Transactions on Quantum Engineering, vol. 2, pp. 1-11, 2021, no. 6003811, https://doi.org/10.1109/TQE.2021.3116540.

[12] N. Messaoudi, C. Crocker and M. Almendros, "A Hardware-Accelerated Qubit Control System for Quantum Information Processing," 2020 XXXV Conference on Design of Circuits and Integrated Systems (DCIS), Segovia, Spain, 2020, pp. 1-5, https://doi.org/10.1109/DCIS51330.2020.9268643

[13] https://www.realdigital.org/hardware/rfsoc-4x2

[14] R. Gebauer, N. Karcher and O. Sander, "A modular RFSoC-based approach to interface superconducting quantum bits," 2021 International Conference on Field-Programmable Technology (ICFPT), Auckland, New Zealand, 2021, pp. 1-9, https://doi.org/10.1109/ICFPT52863.2021.9609909.

[15] X. Guo and M. Schulz, "A Scalable and Cross-Technology Quantum Control Processor," 2023 33rd International Conference on Field-Programmable Logic and Applications (FPL), Gothenburg, Sweden, 2023, pp. 353-354, https://doi.org/10.1109/FPL60245.2023.00063.

[16] M. Sohaib Alam, et al., "Quantum Computing Hardware for HEP Algorithms and Sensing", Proceedings of the US Community Study on the Future of Particle Physics (Snowmass 2021), 29 Apr. 2022, [online] https://arxiv.org/pdf/2204.08605.

[17] E. T. Jaynes and F. W. Cummings, "Comparison of quantum and semiclassical radiation theories with application to the beam maser," in Proceedings of the IEEE, vol. 51, no. 1, pp. 89-109, Jan. 1963, https://doi.org/10.1109/PROC.1963.1664.

[18] Hiroo Azuma, Quantum Computation with the Jaynes-Cummings Model, Progress of Theoretical Physics, Volume 126, Issue 3, September 2011, Pages 369–385, https://doi.org/10.1143/PTP.126.369.

[19] J. Larson, T. Mavrogordatos, "Jaynes-Cummings Model and Its Descendants: Modern Research Directions" IOP Series in Quantum Technology, Bristol, 2021, (IOP Publishing Ltd), pp. 167-179.

[20] D. Niemietz, P. Farrera, S. Langenfeld, G. Rempe, "Nondestructive detection of photonic qubits". Nature 591, 570–574 (2021). https://doi.org/10.1038/s41586-021-03290-z.

[21] R. Lescanne, et al. "Irreversible qubit-photon coupling for the detection of itinerant microwave photons." Physical Review X, vol. 10, no. 2, 18 May 2020, https://doi.org/10.1103/physrevx.10.021038.

[22] H. Johnson, S. Zorzetti and J. Saniie, "Exploration of Optimizing FPGA-based Qubit Controller for Experiments on Superconducting Quantum Computing Hardware," 2023 IEEE International Conference on Electro Information Technology (eIT), Romeoville, IL, USA, 2023, pp. 406-411, https://doi.org/10.1109/eIT57321.2023.10187252.

[23] S. Kwon, et al. "Gate-based superconducting quantum computing." Journal of Applied Physics, vol. 129, no. 4, 28 Jan. 2021, https://doi.org/10.1063/5.0029735.